%Paper: hep-th/9309071
%From: Michael Lashkevich <lashkevi@cpd.landau.free.msk.su>
%Date: Mon, 13 Sep 93 13:40:37 +0400 (MSD)
%Date (revised): Tue, 12 Oct 93 19:55:38 +0400 (MSD)

\font\petit=cmr8  %if you have not cmr8 use cmr10
\font\ow=msbm10   %if you have not msbm10 replace this line for \let\ow=\bf
\def\slq#1{U_{#1}(sl(2))}
\def\pno{\par\noindent}
\def\ref#1{$^{#1}$}
\def\2pi{2\pi i}
\def\M#1{{\rm M}_{#1}}
\def\<{\langle}
\def\>{\rangle}
\def\half{{1\over2}}
\def\quarter{{1\over4}}
%%%%%%%%%%%%%%%%%%%%%%%%%%%%%%%%%%%%%%%%%%%%%%%%%%%%%%%%%%%%%%%%%%%%%%%%%%%%%
\normalbaselineskip = 12 pt
\magnification = 1200
\hsize = 15 truecm \vsize = 22 truecm \hoffset = 1.0 truecm
%\def\headline{\pageno}
%\nopagenumbers
%%%%%%%%%%%%%%%%%%%%%%%%%%%%%%%%%%%%%%%%%%%%%%%%%%%%%%%%%%%%%%%%%%%%%%%%%%%%%
\rightline{LANDAU-93-TMP-6}
\rightline{hep-th/9309071}
\rightline{September 1993}
\rightline{Submitted to Mod. Phys. Lett. A}
\vskip 2 truecm
\centerline{{\bf NEW CONFORMAL MODELS WITH $c<2/5$}}
\vskip 1 truecm
\centerline{M.~YU.~LASHKEVICH\footnote{$^*$}{E-mail:
lashkevi@cpd.landau.free.msk.su}}
\centerline{Landau Institute for Theoretical Physics,}
\centerline{Kosygina 2, GSP-1, 117940 Moscow V-334, Russia}
\vskip 2 truecm
\centerline{ABSTRACT}
\pno
{\petit
The zoo of two-dimensional conformal models has been supplemented by a series
of nonunitary conformal models obtained by cosetting minimal models.
Some of them coincide with minimal models, some do not have
even Kac spectrum of conformal dimensions.
}
%%%%%%%%%%%%%%%%%%%%%%%%%%%%%%%%%%%%%%%%%%%%%%%%%%%%%%%%%%%%%%%%%%%%%%%%%%%%%
\vfill\eject
%%%%%%%%%%%%%%%%%%%%%%%%%%%%%%%%%%%%%%%%%%%%%%%%%%%%%%%%%%%%%%%%%%%%%%%%%%%%%
\pno
In this paper we continue to explore coset constructions of minimal
models.\ref{1,2} Let us designate as $\M{PQ}$ the minimal model with
the central charge of the Virasoro algebra\ref{3}
$$
c_{P,Q}=1-6{(Q-P)^2\over PQ}.
$$
Recall that the monodromy properties of $\M{PQ}$ are described by
braiding irreducible representations of the quantum group
$\slq{q(P,Q)}\times\slq{q(Q,P)}$, where
$$
q(P,Q)=\exp\left(\2pi{Q\over P}\right).
\eqno(1)
$$
The minimal model $\M{PQ}$ is described by vertex operators\ref{4}
$$
\eqalign{
\phi^{\ mn}_{(p,q)}
&(z):{\cal H}_{(p_1,q_1)}\longrightarrow{\cal H}_{(p_1+p-1-2m,q_1+q-1-2n)},
\cr
p
&=1,2,\dots,P-1,\quad q=1,2,\dots,Q-1,
\cr
m
&=0,1,\dots,p-1,\quad\,n=0,1,\dots,q-1,
}
$$
with conformal dimensions
$$
\Delta_{(p,q)}={(Qp-Pq)^2-(Q-P)^2\over 4PQ}.
$$
Here ${\cal H}_{(p,q)}$ is an irreducible representation of the Virasoro
algebra over the state $\phi_{(p,q)}(0)|vacuum\>$,
${\cal H}_{Q-p,P-q}\sim{\cal H}_{(p,q)}$. In the bosonic
representation\ref{5-7,4} the indices $m$ and $n$ mean numbers of screenings.
In terms of quantum group, the pairs $\left(\half(p-1),\half(p-1)-m\right)$
and $\left(\half(q-1),\half(q-1)-n\right)$ are pairs ({\it highest we\-ight,
we\-ight}) or ({\it "moment", "projection of moment"})
of the representation of respective $\slq{x}$ quantum group.

Monodromy invariant fields can be constructed as\ref{6,4,8}
$$
\phi_{(p,q)}(z,\overline{z})
=\sum_{m,n}X_p(m;q(P,Q))X_q(n;q(Q,P))\phi^{\ mn}_{(p,q)}(z)
\overline{\phi^{\ mn}_{(p,q)}(z)},
$$
where coefficients $X_p(m,x)$ are expressed in terms of braiding
matrices of conformal blocks\ref{6} or $R$-matrix of the quantum
group.\ref{8,9}

Consider two models $\M{PS}$ and $\M{SQ}$ with vertices
$\phi^{(1)}{}^{\ mr}_{(p,s)}(z)$ and $\phi^{(2)}{}^{\ r\,n}_{(s,q)}(z)$
respectively. If
$$
q(S,P)=\overline{q(S,Q)},
\eqno(2)
$$
we can consider a convolution of two models\ref{10,1,2} $\M{PS}\M{SQ}$
generated by vertices
$$
\phi_{(}{}^m_{p,s,}{}^n_{q)}(z)
=\sum_rX_s(r;q(S,P))\phi^{(1)}{}^{\ mr}_{(p,s)}(z)
\phi^{(2)}{}^{\ r\,n}_{(s,q)}(z).
\eqno(3)
$$
We shall designate them as
$$
\phi_{(p,s,q)}(z)=\phi^{(1)}_{(p,s)}(z)\phi^{(2)}_{(s,q)}(z)
$$
and call them convolutions of vertex operators. Monodromy
properties of such convolutions are described by the quantum group
$\slq{q(P,S)}\times\slq{q(Q,S)}$. The multipliers $\slq{q(S,P)}$
and $\slq{q(S,Q)}$ connected to indices $s$ and $r$ drop out.

Conditon (2) holds, if
$$
P+Q=NS,\quad N\in{\ow Z}.
\eqno(4)
$$
If we want to consider a coset construction $\M{PS}\M{SQ}/$({\it something}),
we must construct the energy-momentum tensor of the denominator
in terms of fields of the numerator. The vertices $\phi_{(1,s,1)}(z)$
possess trivial monodromy properties and can be considered as chiral
currents.
Thus, we shall look for the energy-momentum
tensor of the denominator, $T_H(z)$, and that of the coset construction,
$T_C(z)$, in the form
$$
\eqalign{
T_H(z)
&=\phantom{(1\ -\ }A\phantom{)}T_1(z)+\phantom{(1\ -\ }B\phantom{)}T_2(z)
+C\phi_{(1,s_0,1)}(z),
\cr
T_C(z)
&=(1-A)T_1(z)+(1-B)T_2(z)-C\phi_{(1,s_0,1)}(z),
}\eqno(5)
$$
where $A$, $B$ and $C$ are constants, $T_1(z)$ and $T_2(z)$ are the
energy-momentum tensors of $\M{PS}$ and $\M{SQ}$ respectively.
The third term in (5) must be of conformal dimension 2:
$$
\Delta^{(1)}_{(1,s_0)}+\Delta^{(2)}_{(s_0,1)}
\equiv{s_0-1\over 4S}[(P+Q)(s_0+1)-4S]=2.
\eqno(6)
$$
Both conditions (4) and (6) are satisfied only for
$s_0=2$, $N=4$ and $s_0=3$, $N=2$. The case $s_0=3$, $N=2$ for
unitary models was considered earlier,\ref{1,2} and its generalization
to nonunitary models is nearly straightforward. In this paper we
shall concentrate on the other case
$$
s_0=2,\quad P+Q=4S.
\eqno(7)
$$

Using\hfil bosonic\hfil representation\hfil we\hfil obtain\hfil
the\hfil operator\hfil product
\hfil expansion (OPE) for the chiral current $\phi_{(1,2,1)}(z)$
$$
\eqalign{
\phi_{(1,2,1)}(z')\phi_{(1,2,1)}(z)
&={1\over(z'-z)^4}+{2\theta(z)\over(z'-z)^2}
+{\partial\theta(z)\over z'-z}+O(1),
\cr
\theta(z)
&={2P\over 3Q-5P}\,T_1(z)+{2Q\over 3P-5Q}\,T_2(z),
}\eqno(8)
$$
where $\partial\equiv\partial/\partial z$, $O(1)$ designates the
terms regular at $z'\longrightarrow z$. Now it is easy to check that
the currents
$$
\eqalign{
T_H(z)
&=-{2\over5}{P\over Q-P}\,T_1(z)+{2\over5}{Q\over Q-P}\,T_2(z)
\cr
&+i{\sqrt{2(3Q-5P)(3P-5Q)}\over 5(Q-P)}\ \phi_{(1,2,1)}(z),
\cr
T_C(z)
&={1\over5}{5Q-3P\over Q-P}\,T_1(z)+{1\over5}{3Q-5P\over Q-P}\,T_2(z)
\cr
&-i{\sqrt{2(3Q-5P)(3P-5Q)}\over 5(Q-P)}\ \phi_{(1,2,1)}(z)
}\eqno(9)
$$
obey the OPE's
$$
\eqalign{
T_i(z')T_i(z)
&={\half c_i\over(z'-z)^4}+{2T_i(z)\over(z'-z)^2}
+{\partial T_i(z)\over z'-z}+O(1),\quad i=H,C,
\cr
T_H(z')T_C(z)
&=O(1),
}$$
where the central charges are given by
$$
\eqalign{
c_H
&=-{22\over5},
\cr
c_C
&={(3Q-5P)(3P-5Q)\over 10PQ}<{2\over5}.
}\eqno(10)
$$
$c_H$ is the central charge of the minimal model $\M{2,5}$. Thus,
we shall consider the coset construction
$$
{\M{PS}\M{SQ}\over\M{2,5}},\quad S={P+Q\over4}\in{\ow Z}.
\eqno(11)
$$

Now we direct our attention to primary fields of the coset model.
Consider the OPE
$$
\eqalign{
\phi_{(1,2,1)}(z')\phi_{(p,s,q)}(z)
&\sim(z'-z)^{-1-2\left(s-{p+q\over4}\right)}\left[\phi_{(p,s-1,q)}\right]
\cr
&+(z'-z)^{-1+2\left(s-{p+q\over4}\right)}\left[\phi_{(p,s+1,q)}\right].
}\eqno(12)
$$
We write down clearly the factors of the kind $(z'-z)^\alpha$ at the
fields of the lowest dimensions in conformal families. If
$$
\quarter(p+q-2)\leq s\leq\quarter(p+q+2),
$$
there are no poles of the power $>2$ in the expansion (12), and the field
$\phi_{(p,s,q)}$ can be primary with respect to the coset energy-momentum
tensor $T_C(z)$ from (9).

We shall discuss all cases in sequence.

1. $p+q\in4{\ow Z}$, $s=\quarter(p+q)$. In this case
$$
\eqalign{
T_C(z')\phi_{(p,s,q)}(z)
&\sim(z'-z)^{-2}\left[\phi_{(p,s,q)}\right]
\cr
&+(z'-z)^{-1}\left(\left[\phi_{(p,s-1,q)}\right]
+\left[\phi_{(p,s+1,q)}\right]\right).
}$$
The conformal dimension of the field $\phi_{(p,s,q)}$ with respect to
$T_C(z)$ is given by
$$
\Delta^0_{p,q}={(Qp-Pq)^2-(Q-P)^2\over16PQ}-{1\over20},
\eqno(13)
$$
and the conformal dimension with respect to $T_H(z)$ is $-{1\over5}$.
It means that
$$
\phi'_{(1,2)}(z)\phi^0_{p,q}(z)=\phi^{(1)}_{(p,s)}(z)\phi^{(2)}_{(s,q)}(z),
\quad p+q\in4{\ow Z},\quad s=\quarter(p+q),
\eqno(14)
$$
where $\phi'_{(1,2)}(z)$ is the primary field of the conformal
dimension $-{1\over5}$ in the model $\M{2,5}$, and $\phi^0_{p,q}(z)$
are vertices of the coset model (11). There is a convolution of
$\phi'_{(1,2)}(z)$ and $\phi^0_{p,q}(z)$ in the left-hand side
of (14). Monodromy properties of the coset model are described
by the quantum group $\slq{q(P,S)}\times\slq{q(S,Q)}\times
\slq{\overline{q(2,5)}}$.

2. $p+q\pm1\in4{\ow Z}$, $s=\quarter(p+q\pm1)$. In this case
$$
\phi_{(1,2,1)}(z')\phi_{(p,s,q)}\sim(z'-z)^{-{3\over2}}\cdot(something).
$$
Therefore, the product $T_C(z')\phi_{(p,s,q)}(z)$ contains in its
decomposition half-integer powers of $(z'-z)$ as well as integer ones.
It means that $T_C(z)$ is no longer a chiral current.
Fortunately, one can eliminate this sector, because there are no
fields $\phi_{(p,s,q)}$ with odd $p+q$ in fusions of fields with
even $p+q$.

3. $p+q\pm2\in4{\ow Z}$, $s_\pm=\quarter(p+q\pm2)$, $s_+-s_-=1$.
The fields $\phi_{(p,s_+,q)}(z)$ and $\phi_{(p,s_-,q)}(z)$ have the same
conformal dimensions with respect to $T_1(z)+T_2(z)$. In other words,
$$
\eqalign{
\phi_{(1,2,1)}(z')\phi_{(p,s_+,q)}(z)
\sim(z'-z)^{-2}\left[\phi_{(p,s_-,q)}\right]+O(1),
\cr
\phi_{(1,2,1)}(z')\phi_{(p,s_-,q)}(z)
\sim(z'-z)^{-2}\left[\phi_{(p,s_+,q)}\right]+O(1).
}$$
The operator $L^C_0=\oint{du\over\2pi}(u-z)T_C(u)$ mixes fields
$\phi_{(p,s_+,q)}(z)$ and $\phi_{(p,s_-,q)}(z)$. Conformal dimensions
in the coset model are eigenvalues of this operator. Diagonalizing it
we obtain two fields
$$
\eqalignno{
\phi^-_{p,q}(z)
&=\sqrt{y+\half}\ \phi^{(1)}_{(p,s_+)}(z)\,\phi^{(2)}_{(s_+,q)}(z)
+i\sqrt{y-\half}\ \phi^{(1)}_{(p,s_-)}(z)\,\phi^{(2)}_{(s_-,q)}(z),
&\cr
&&(15{\rm a})
\cr
\phi'_{(1,2)}(z)\,\phi^+_{p,q}(z)
&=-i\sqrt{y-\half}\ \phi^{(1)}_{(p,s_+)}(z)\,\phi^{(2)}_{(s_+,q)}(z)
+\sqrt{y+\half}\ \phi^{(1)}_{(p,s_-)}(z)\,\phi^{(2)}_{(s_-,q)}(z),
&\cr
&&(15{\rm b})
\cr
y
&={Qp-Pq\over2(Q-P)},\quad p+q-2\in4{\ow Z},\quad s_\pm=\half(p+q\pm2)
&(15{\rm c})}
$$
with conformal dimensions
$$
\eqalignno{
\Delta^-_{p,q}
&={(Qp-Pq)^2-(Q-P)^2\over16PQ},
&(16{\rm a})
\cr
\Delta^+_{p,q}
&={(Qp-Pq)^2-(Q-P)^2\over16PQ}+{1\over5}.
&(16{\rm b})}
$$

Other primary fields can appear in such models too, but at present there
is no simple method to find them.

Consider some examples. The first example is $\M{2,3}\M{3,10}/\M{2,5}$.
The central charge $c_C=-22/5$ coincides with that of the minimal model
$\M{2,5}$. The conformal dimensions of the coset primary fields
$$
\Delta^-_{1,1}=\Delta^-_{1,9}=\Delta^+_{1,5}=0,
\quad\Delta^0_{1,3}=\Delta^0_{1,7}=\Delta^-_{1,5}=-{1\over5}
$$
confirm the identification
$$
{\M{2,3}\M{3,10}\over\M{2,5}}\sim\M{2,5}.
$$

For $\M{2,5}\M{5,18}/\M{2,5}$, $c=-154/15$, the conformal dimensions
are given by
$$
\eqalign{
&\Delta^-_{1,1}=0,\quad\Delta^0_{1,3}=\Delta^+_{1,9}=-{11\over45},
\quad\Delta^-_{1,5}=-{1\over3},
\cr
&\Delta^+_{1,5}=-{2\over15},\quad\Delta^0_{1,7}=-{7\over15},
\quad\Delta^-_{1,9}=-{4\over9}.
}$$
We can identify this model at least with some sector in $\M{5,18}$.

For $\M{5,4}\M{4,11}/\M{2,5}$ the central charge $c=-32/55$ corresponds
to an irrational conformal model. The conformal dimensions
$$
\eqalign{
&\Delta^-_{1,1}=0,\quad\Delta^0_{1,3}=-{4\over55},
\quad\Delta^0_{2,2}={4\over55},\quad\Delta^0_{3,1}={4\over5},
\cr
&\Delta^-_{1,5}={2\over11},\quad\Delta^-_{2,4}=-{2\over55},
\quad\Delta^-_{3,3}={18\over55},\quad\Delta^-_{4,2}={14\over11},
\cr
&\Delta^+_{1,5}={21\over55},\quad\Delta^+_{2,4}={9\over55},
\quad\Delta^+_{3,3}={29\over55},\quad\Delta^+_{4,2}={81\over55},
\cr
&\Delta^0_{1,7}={31\over55},\quad\Delta^0_{2,6}=-{1\over55},
}$$
do not generally coincide with any Kac conformal dimensions.
\medskip\pno
{\bf Acknolegements}
\medskip

This work was supported in part by the Landau Scholarship Grant
awarded by Forschungszentrum J\"ulich and the Soros Foundation Grant
awarded by the American Physical Society.
%%%%%%%%%%%%%%%%%%%%%%%%%%%%%%%%%%%%%%%%%%%%%%%%%%%%%%%%%%%%%%%%%%%%%%%%%%%%%%%
\vfill\eject
%%%%%%%%%%%%%%%%%%%%%%%%%%%%%%%%%%%%%%%%%%%%%%%%%%%%%%%%%%%%%%%%%%%%%%%%%%%%%%%
\pno
{\bf References}
\medskip\pno
\halign{\hfill#&#\hfill\cr
1. & M. Yu. Lashkevich, {\it Mod. Phys. Lett.} {\bf A8}, 851 (1993)
\cr
2. & M. Yu. Lashkevich, preprint LANDAU-93-TMP-3,
hep-th/9304116
\cr
&(Apr. 1993); to be published in {\it Int. J. Mod. Phys.}
\cr
3. & A. A. Belavin, A. M. Polyakov and A. B. Zamolodchikov,
{\it Nucl. Phys.} {\bf B241}, 333 (1984)
\cr
4. & G. Felder, {\it Nucl. Phys.} {\bf B317}, 215 (1989)
\cr
5. & Vl. S. Dotsenko and V. A. Fateev, {\it Nucl. Phys.}
{\bf B240 [FS12]}, 312 (1984)
\cr
6. & Vl. S. Dotsenko and V. A. Fateev, {\it Nucl. Phys.}
{\bf B251 [FS13]}, 691 (1985)
\cr
7. & Vl. S. Dotsenko and V. A. Fateev, {\it Phys. Lett.}
{\bf B154}, 291 (1985)
\cr
8. & G. Felder, J. Fr\"olich and G. Keller,
{\it Commun. Math. Phys.} {\bf 130}, 1 (1990)
\cr
9. & C. Gomez and G. Sierra, {\it Nucl. Phys.}
{\bf B352}, 791 (1991)
\cr
10. & M. B. Halpern and N. Obers, preprint LBI-32619, USB-PTH-92-24,
\cr
&BONN-HE-92/21, hep-th/9207071 (July 1992)\cr}
\end